\documentclass[aps,prb,twocolumn,superscriptaddress]{revtex4}

\usepackage{amssymb}
\usepackage{amsmath}
\usepackage{amsfonts}
\usepackage{graphicx}% Include figure files
\usepackage{bm}% bold math

\begin{document}
%\preprint{APS/123-QED}
%\parindent=0.5cm
\title{Stability of a skyrmion and interaction of magnons}
\author{D.N.Aristov}
\affiliation{Department of Physics, St.Petersburg State University, Ulianovskaya 1, St.Petersburg 198504, Russia}
\affiliation{ ``PNPI'' NRC ``Kurchatov Institute'', Gatchina 188300, Russia}
\author{P.G.Matveeva }
\affiliation{Department of Physics, St.Petersburg State University, Ulianovskaya 1, St.Petersburg 198504, Russia}
\affiliation{ ``PNPI'' NRC ``Kurchatov Institute'', Gatchina 188300, Russia}
\date{\today}       

\begin{abstract}
The stability of a single Belavin-Polyakov (BP) skyrmion in isotropic Heisenberg ferromagnet is studied. Such  skyrmion is higher in energy than the uniform ferromagnetic state and is thus metastable. Starting from the lattice model in two spatial dimensions and using Maleyev-Dyson representation for spin operators, we examine the effects of magnon-magnon interaction for two quantities at $T=0$. First we discuss the self-energy corrections to magnon energy. Second we analyze the two-particle Green's function and possible bound states of two magnons.  The simplicity of the model makes possible full analytic treatment of all relevant processes. 
We found that the magnons remain well-defined quasiparticles with a finite lifetime.  The bound states of two magnons are suppressed near the skyrmion,   although they are not excluded far away from it. 
A resonance for the magnons of dilational mode in the vicinity of BP skyrmion is also found, which leads to a redistribution of the spectral weight. 
We conclude that the BP skyrmion as the classical topological object is not  destroyed by quantum fluctuations.
\end{abstract}

%\pacs{75.10.Jm, 75.30.Ds, 75.70.Kw} 

\maketitle

%\tableofcontents

\section{\label{sec:level1}Introduction}

Topological defects in magnets is a topic of large interest today. Understanding their behaviour in external fields, their stability and other properties are important for developing and creating new types of memory\cite{Fert2013,Koshibae2015,Romming2013}. Examples of such defects include domain walls and peculiar twisted magnetic structures called skyrmions. These skyrmions are topological objects, which means that they cannot be transformed to the trivial collinear magnetic order in a continuous fashion. 

The pioneering work of Belavin and Polyakov (BP)  \cite{Belavin1975}  discussed skyrmions in ferromagnets with isotropic Heisenberg interaction.  The proposed BP skyrmion solution provided a local minimum of the classical energy, but this energy was higher than for the uniformly magnetized  state.  This means that  the BP skyrmion is a metastable solution, but is not unstable one because the spectrum of small fluctuations is non-negative. 

A number of theoretical works suggested various ways to lower the classical energy of the skyrmion state and thus make it energetically favorable  \cite{Bogdanov1994,Ezawa2010,  Mochizuki2012, Han2010, Nagaosa2013} (see also \cite{Aristov2015} and references therein). This process of adding weaker interactions to the main Heisenberg Hamiltonian  is called stabilization.  In the context of condensed matter it was found both theoretically \cite{Bogdanov1994} and  experimentally  \cite{Yu2010} that such stabilization can be perhaps most simply achieved by simultaneous presence of Dzyaloshinskii-Moriya interaction and the uniform magnetic field. For the fields in the appropriate range  the skyrmion solution delivers the minimum to classical energy of the magnet. More precisely, the skyrmion ground state assumes the finite density of skyrmions, eventually forming the skyrmion superlattice or ``skyrmion crystal''. \cite{Iwasaki2014,Lin2014,Schutte2014,Tatara2014,Kovrizhin2013}  Small fluctuations of localized moments around the skyrmion (static) ground state describe the magnon dynamics which can be studied, e.g. by corresponding Landau-Lifshitz equation \cite{Lin2014} or within semiclassical linear spin-wave theory. \cite{Schutte2014} 

The quantum corrections to the obtained dynamical quantities are usually ignored by theorists, which may generally be justified by several arguments. First, the corrections correspond to the interaction of magnons, which is formally small in semiclassics by the inverse value of localized moment, $1/s$. Second, it is well known that for the uniformly magnetized ferrromagnetic state the corrections to magnon energy, $\epsilon_{q}$, vanish for zero temperature, so that magnons are ideal quasiparticles. Third, the appearance of ``stabilizing''  additional interactions makes the analytical approach to the problem very hard:  the numerics should be used already for determination of spectrum and wave functions, and  further calculation of corrections becomes nearly impossible.  

First two reasons to discard quantum corrections become questionable for $s\sim1$ and for non-uniform skyrmion static solution. The third reason of intractability is weakened in case of BP skyrmion for the isotropic Heisenberg interaction, where explicit analytical formulas are available.    The consideration of the latter case provides a possibility to question the instability of  topologically protected objects.   The instability implies a spontaneous transition to the true ground state, whereas topological character of the metastable skyrmion prevents it from destruction by fluctuations. Such destruction is different from the recently studied case  \cite{Verga2014} of magnetic dynamics with dissipation.

In the present paper we study the effects of interaction between magnons in the presence of one BP skyrmion in two spatial dimensions and at zero temperature. This slightly artificial case allows us to fully employ analytical approach in the  semiclassical $1/s$ expansion and explicitly find the leading contributions to the self-energy corrections and the two-particle magnon Green's function.  The additional interest in the case of one BP skyrmion is related to the presence of three zero modes  associated with three broken conformal symmetries of the problem \cite{Ivanov1989,Ivanov1995,Ivanov1999,Aristov2015}, see also \cite{Buijnsters2014}.  Any second-order quantum correction to the energy is non-positive and a strictly negative correction to zero modes would make the system unstable.   

Ultimately we find that the corrections  are parametrically small in low-energy sector and exactly vanish for zero modes, which means that magnons remain well-defined excitations in the presence of BP skyrmion.  It means that although  the skyrmion  does not provide a global minimum in energy, the small fluctuations around this local minimum are long-lived. Two particle Green's function for the usual uniform ferromagnet is known to contain poles, corresponding to the bound states of two magnons. \cite{Wortis1963}  Our analysis below shows that the corresponding quantity, the dressed four-vertex of interaction, is suppressed around BP skyrmion. The bound states may in principle be realized for small energies, but their wave-functions avoid the vicinity of the skyrmion.  Our full non-perturbative treatment of infrared modes indicates that the BP skyrmion is the stable formation. The probability of transition to the true uniform ground state  is apparently strictly zero at $T=0$.  

This said, we should also point out that the absence of true bound state in the vicinity of the BP skyrmion is accompanied by the appearance of the resonance in the 4-vertex of scattering of two magnons, related to soft dilational mode.  This resonance lies in the continuum spectrum, is well defined in the limit of large $s$ and leads to redistribution of the spectral weight in the Green's function for corresponding magnons. The resonance disappears in three dimensions and we provide a criterion for this crossover.

  %%%%%%%%%%%%%%%%%%%%%%%%%%%%%%
 %
 %%%%%%%%%%%%%%%%%%%%%%%%%%%%%%

  The rest of the paper is organized as follows. We formulate our model, and found an explicit form of interaction between magnons in the presence of one BP skyrmion in Section \ref{sec:level2}. Focusing on  the most interesting case of magnons, whose magnetic quantum number corresponds to zero modes, we discuss 
the self-energy corrections and the magnons' lifetime  in Section  \ref{sec:level3}. The two-particle Green's function is studied in Section \ref{sec:level4}, where the complete asymptotic solution of the Bethe-Salpeter equation for the dressed interaction is found.  The generalization to the three-dimensional case is briefly discussed here. We present our conclusions in Section  \ref{sec:level6}. The density of magnon states is discussed in Appendix   \ref{app}. 

\section{\label{sec:level2}Magnon interaction in presence of the skyrmion}

We start with the lattice Heisenberg exchange Hamiltonian:
\begin{equation}
H=\sum_{i,j}J(\mathbf{r}_i-\mathbf{r}_j)\mathbf{S}_{\mathbf{r}_i}\mathbf{S}_{\mathbf{r}_j}
\label{Ham}
\end{equation} in two spatial dimensions (2D). 
It is well known that the quantum fluctuations destroy the long-range order in 2D at any non-zero temperature, and we consider only the case $T=0$. The non-interacting magnon Hamiltonian below \eqref{LSWT}  does not contain anomalous terms (cf.\  \cite{Schutte2014}), so that zero-point motion of spins is absent and the average spin value has its maximum value, $| \langle\mathbf{S} \rangle | =s$.

We assume that the normalized magnetization density $\mathbf{n}=\mathbf{S}/s=(\sin\beta \cos\alpha,\sin\beta \sin\alpha ,\cos\beta$ is non-uniform and corresponds to Belavin-Polyakov skyrmion state \cite{Belavin1975}, explicitly  \begin{equation}
 \label{BPsol}
 \beta=2\arctan\left(\frac{r}{r_0}\right), \alpha+\alpha_0=\phi
 \end{equation} 
where $r_0$ skyrmion radius and we use the polar coordinates $(r,\phi)$ with the origin at skyrmion's center. Such configuration is characterized by  topological charge  $Q=1$, with 
  \begin{equation}
\label{defQ}
Q=\frac{1}{8\pi} \int d^2 \mathbf{r} \epsilon_{\mu\nu}\epsilon_{abc} n^a\nabla_{\mu}n^b\nabla_{\nu}n^c
\end{equation}
 
In order to discuss the interaction of magnons below, we sketch the derivation of magnon Hamiltonian, starting from the lattice model. \cite{Aristov2015} The lattice model with localized moments allows the well-known Maleyev-Dyson representation for spin operators, preserving the spin commutation relations, $[ {S}^a, {S}^b]=i\epsilon_{abc} {S}^c$:
\begin{equation}
\begin{aligned}
 S^{x}_i & =\sqrt{\frac{s}{2}}\left(a^{\dagger}_i+a_i-\frac{a^{\dagger}_ia^{\dagger}_ia_i}{2s}\right) \\
 S^{y}_i& =-i\sqrt{\frac{s}{2}}\left(-a^{\dagger}_i+a_i+\frac{a^{\dagger}_ia^{\dagger}_ia_i}{2s}\right)  \\
 S^{z}_i& =s-a^{\dagger}_i a_i
 \end{aligned}  
  \label{DM-rep}
\end{equation}
with $[a_j, a_j^{\dagger}]=1$ and the semiclassical limit $s \gg 1$ implied. 
Eq. \eqref{DM-rep} assumes that the  magnetization $\langle \tilde{\mathbf{S}}_{\mathbf{r}}\rangle  $ is directed  along the local $\hat{z}$-axis and it  is convenient to rewrite the Hamiltonian (\ref{Ham}) in  such basis.  The transition to this local basis, $\mathbf{S_r}=\hat{U}(\mathbf{r})\tilde{\mathbf{S}}_{\mathbf{r}}$, is given by the position-dependent  matrix 
\begin{equation}
\label{Urot}
\hat{U}(\mathbf{r})=e^{-\alpha\sigma_3}e^{-\beta\sigma_2}e^{-\gamma\sigma_3}
\end{equation}
with $\sigma_3, \sigma_2$ generators of SO(3) group, and $\alpha,\beta, \gamma$ Euler angles. The Hamiltonian (\ref{Ham}) takes then the form: 
\begin{equation}
\label{Hrot}
H=\sum_{\mathbf{r,n}}J(\mathbf{n})\tilde{\mathbf{S}}_{\mathbf{r}}\hat{R}(\mathbf{r,n})\tilde {\mathbf{S}}_{\mathbf{r+n}}
\end{equation}
with  $\hat{R}(\mathbf{r,n})=\hat{U}^{-1}(\mathbf{r})\hat{U}\mathbf{(r+n)}$. 
We assume rapid decrease of $J(\mathbf{n})$ with distance $\mathbf{n}$, and slow variation of $\hat{U}(\mathbf{r})$ on the scale of lattice spacing, $a$. We expand the matrix $\hat{R}(\mathbf{r,n})$ in a Taylor series upon $\mathbf{n}$ and perform the variation of the classical energy on $\alpha, \beta, \gamma$. This is done by putting $a,a^{\dagger}$ to zero in \eqref{DM-rep}.  First non-trivial  local minimum is given by Eq.\ \eqref{BPsol} with $r_{0}\gg a$ implied.  

Knowing the explicit expressions for matrices $\hat{U}(\mathbf{r}), \hat{R}(\mathbf{r,n})$ (see \cite{Aristov2015}) and using  (\ref{DM-rep}) we represent the appearing bosonic Hamiltonian as a formal expansion in  small parameter $1/s$ : 
\begin{equation}
H=s^2  E_{class}+s  H^{(2)}+ \sqrt{s}  H_{int}^{(3)}+s^0  H_{int}^{(4)} +\frac1{\sqrt{s}}%s^{-1/2} 
H_{int}^{(5)}\,,
\label{Ham-expa}
\end{equation}
The first term $E_{class}$ corresponds to the classical energy of the magnet: 
\begin{equation}
s^2 E_{class}= - s^2 J_{0} V + 4 s^{2} C \int d\mathbf{r} \frac{r_0^2}{(r^2+r_0^2)^2}  
\label{Eclass}
\end{equation}
with  $ \sum J(\mathbf{n}) e^{i \mathbf{qn}} \simeq -J_{0}+ C q^{2}$. Here $J_{0} > 0$ and $ C\sim  J_{0} a^2 >0$ . The first term in (\ref{Eclass}) is proportional to crystal volume, $V$, and  gives the energy of uniformly magnetized sample, and the second contribution, $\delta E=4\pi Cs^{2}>0$ shows that the skyrmion configuration is higher in energy and independent of its size $r_0$ and orientation $\alpha_{0}$ in \eqref{BPsol}.  

The second term $H^{(2)}$ in (\ref{Ham-expa}) is quadratic in boson operators and describes the linear spin wave theory (LSWT) in ferromagnet in the presence of a skyrmion. In continuum limit at $r_{0}\gg a$ we have  \cite{Aristov2015}
\begin{equation}
\begin{aligned}
s H^{(2)} &= s C \int d\mathbf{r}  \, a^{\dagger}_{\mathbf{r}} \hat H^{(2)}
a_{\mathbf{r}}   \\
\hat H^{(2)} & =
 -\nabla^{2} + \frac{4  L_{z}}{r^2+r_0^2} +4\frac{r^2-r_0^2}{(r^2+r_0^2)^2} 
\end{aligned}
\label{LSWT} 
\end{equation}
with the appearance of $L_{z} = - i \partial / \partial \phi$ showing the chiral character of the skyrmion. 
The relevant information about the low-energy magnon states is given in Section \ref{sec:level3} and in Appendix. 
In what follows, we measure distances  in units of skyrmion size $r_{0}$ and the energies in quantities  $\epsilon_0=s C r_0^{-2} \ll J_{0}$.

% \section{\label{sec:level3}Self-energy corrections to magnon spectrum} 

The  %cubic $H_{int}^{(3)}$ and quartic $H_{int}^{(4)}$ 
higher -order terms in   (\ref{Ham-expa})   have  the form 
\begin{equation}
\begin{aligned}
 s^{1/2}{H}_{int}^{(3)} & = - \sqrt{\frac2s} \int d\mathbf{r}\, \left ( a_{\mathbf{r}}^{\dagger}a_\mathbf{r}^{\dagger}  \hat{H}^{(3)} a_{\mathbf{r}}  + H.c. \right ) \,, \\
\hat{H}^{(3)} & = \frac{e^{-i\phi}  }{ 1+r^2} \left(  \frac{2  r }{1+r^2 }   +
   \frac{\partial}{\partial r} +  \frac{1}{r}\hat{L}_z\right)  
   \equiv \frac{e^{-i\phi}  }{ 1+r^2} \hat A_{\mathbf{r}}     \,,  
  \\ % \end{aligned}  \label{12} \end{equation}  \begin{equation} \begin{aligned}
 {H}_{int}^{(4)}&=\frac{1}{2s}\int d\mathbf{r} \left( \hat{H} ^{(4)}a_{\mathbf{r}_{1}}^{\dagger} a_{\mathbf{r}} ^{\dagger}\right) _{\mathbf{r}_{1}\to \mathbf{r}} \cdot
 a_{\mathbf{r}}a_{\mathbf{r}} \,,\\
\hat{H} ^{(4)} & =\frac{4 \left(1-r^2\right)}{(r^2+1)^2}+ \frac{2\hat{L}_z}{r^2+1} 
+ \nabla_{\mathbf{r}_{1}}\nabla_{\mathbf{r}} 
\\   & =
\frac{4  }{(r^2+1)^2} + \left( \hat A_{\mathbf{r}} 
 - \frac{4 r }{r^2+1 }  \right) \hat A^{\ast}_{\mathbf{r_{1}}} 
 \, , \\ 
  {H}_{int}^{(5)}&=- \frac{s^{-3/2}}{\sqrt{2}}\int d\mathbf{r} 
  \left(
  \hat{H} ^{(5)}a_{\mathbf{r}}^{\dagger} \right) 
    a_{\mathbf{r}}^{\dagger}a_{\mathbf{r}}^{\dagger} 
    a_{\mathbf{r}} a_{\mathbf{r}}   \,,\\
 \hat{H} ^{(5)} & =\frac{e^{-i\phi}  }{ 1+r^2} 
 \left (  \hat A_{\mathbf{r}} -\frac{4r}{r^2+1 }   \right)
\end{aligned} 
\label{Hint}
\end{equation} 
and the last line for $\hat{H} ^{(4)}$ is obtained in the limit $\mathbf{r}_{1}\to \mathbf{r}$. The term $  {H}_{int}^{(5)}$ is shown here for completeness and is inessential for our analysis below.

%Hereinafter energy $\epsilon$ and $r$ are dimensionless: $\epsilon \rightarrow \epsilon/\epsilon_0$, where $\epsilon_0=C/r_0^2$ and $r\rightarrow r/r_0$. We note that for skyrmionic solution $\gamma=Q\phi$, where $Q$ is integer winding number (or topological charge in general). In our work we consider for simplicity $Q=1$. 

The local magnon operators are given by $
a_{\mathbf{r}}  = \sum_{m,\epsilon} \Psi_{m,\epsilon}(\mathbf{r}) a_{m,\epsilon}
$
 where   $\Psi_{m,\epsilon} (\mathbf{r})= e^{-im\phi} \psi_{m,\epsilon}(r)$ is the eigenfunction of $H^{(2)}$ with  the energy,  $\epsilon$, and the angular momentum, $-m$ .  For our analysis it is convenient to  use the mixed coordinate-frequency representation for Green's function: 
\begin{equation}
 G(\omega,\mathbf{r_1,r_2})= \sum_{m,\epsilon}\frac{\Psi^{\ast}_{m,\epsilon}(\mathbf{r}_1)\Psi_{m,\epsilon}(\mathbf{r}_2)}{\omega-\epsilon+i0} 
 \label{Green}
 \end{equation}
In our notation there are three zero modes with $\epsilon=0$ and $m=0,1,2$ of the form 
\begin{equation}
\psi_{m, 0}= \frac{r^{ m}}{1+r^{2}} 
\label{zeromodes}
\end{equation}
which correspond to three conformal symmetries broken by the skyrmion. As explained in \cite{Aristov2015}, these functions with $m=0,1,2$ correspond to translational symmetry, dilation/rotation of the skyrmion and special conformal symmetry (SCT), respectively.  The breaking of the latter symmetry can be regarded as changing of the direction of spins at infinity, with a simultaneous shift in the skyrmion's position.

 \section{\label{sec:level3} Self-energy correction}

\begin{figure}
(a) \includegraphics[width=0.4\columnwidth]{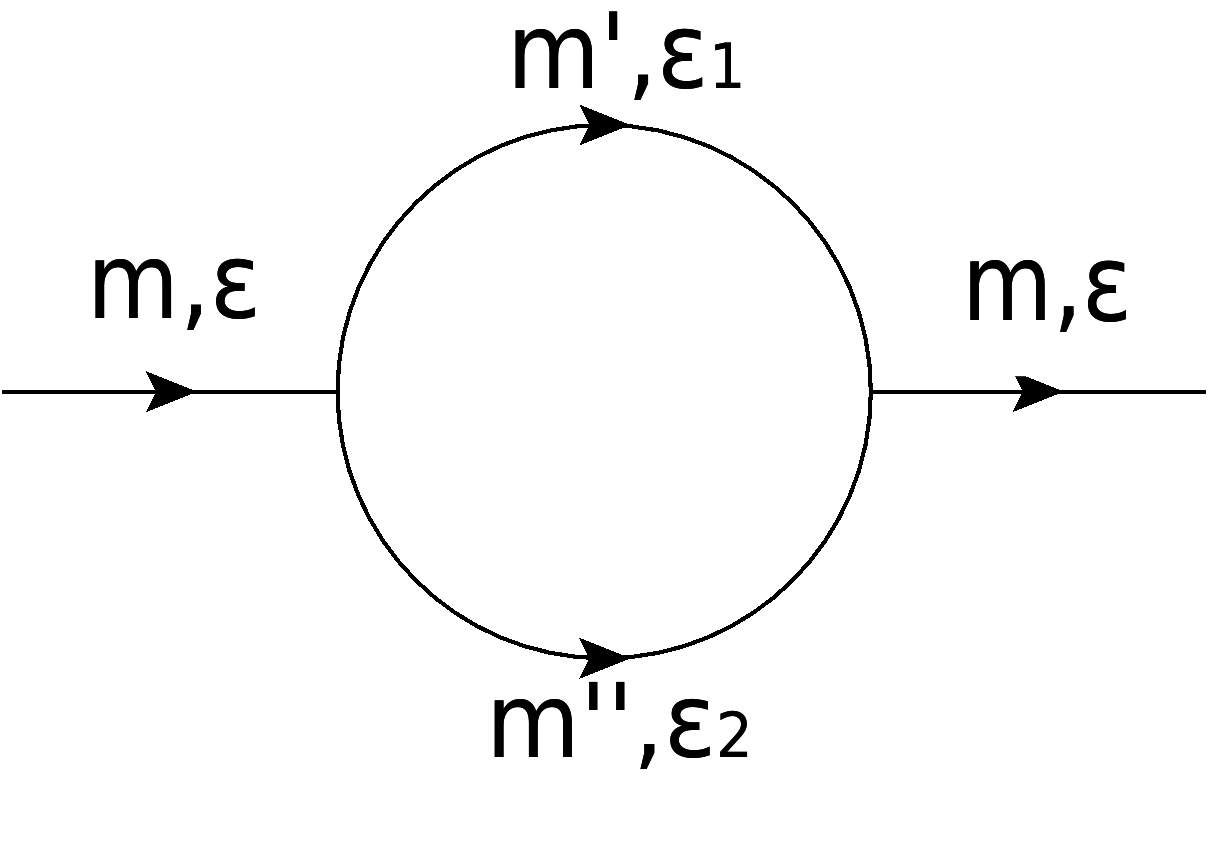} \\
(b) \includegraphics[width=0.4\columnwidth]{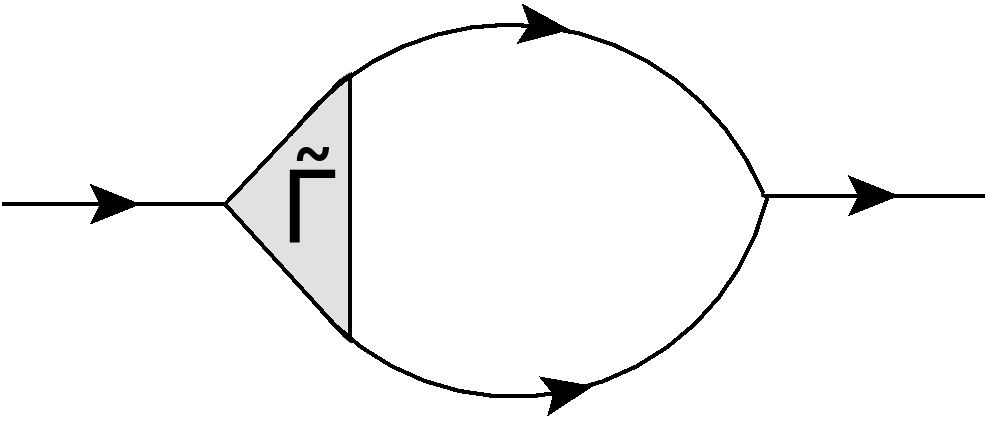}
\caption{\label{SEdiagram} (a) The diagram for the self-energy part in second order of perturbation theory. (b) The self-energy part with the dressed 3-vertex, graphically defined in Fig.\ \ref{fig:vertices} and  
discussed below. }
\end{figure}

The self-energy correction is given by the general diagram shown in Fig. \ref{SEdiagram}.  In the second order of perturbation theory the only contribution surviving in the limit $T=0$ is depicted in  Fig. \ref{SEdiagram}a. This is the correction  of order $1/s$, and in higher orders this main diagram requires both dressing the Green's functions and the dressed 3-vertex, see Fig. \ref{SEdiagram}b. 
As shown below, the dressing of 3-vertex is not important in most cases, but it is essential for the states with $m=1$, corresponding to dynamical dilatation and rotation of the skyrmion. 

In the lowest $1/s$ order the self-energy correction takes the form
 \begin{equation}
 \label{14}
\Sigma(\omega, m, \epsilon)= \sum_{\epsilon_{i} , m_{i}} \frac{|\Gamma(\{m,\epsilon\})|^2 }{ \omega-\epsilon_1-\epsilon_2+i0},
\end{equation}
The vertex amplitude $\Gamma$ is defined by  
  ${H}_{int}^{(3)}$  in    \eqref{Hint} as 
\begin{equation}
\label{15}
\begin{split}
\Gamma(\{m,\epsilon\})=\int d^{2}\mathbf{r} \, \Psi^{*}_{m'\epsilon'}(\mathbf{r})\Psi_{m''\epsilon''}^{*}(\mathbf{r})\hat{H}^{(3)} \Psi_{m\epsilon}(\mathbf{r}), 
\end{split}
\end{equation}
The integration over $\phi$ here  leads to specific selection rules for quantum number $m$, namely $m + 1=m' + m''$.   We show the decay processes involving ``nearly'' zero modes of  Fig.2. 

\begin{figure}
\includegraphics[width=0.7\columnwidth]{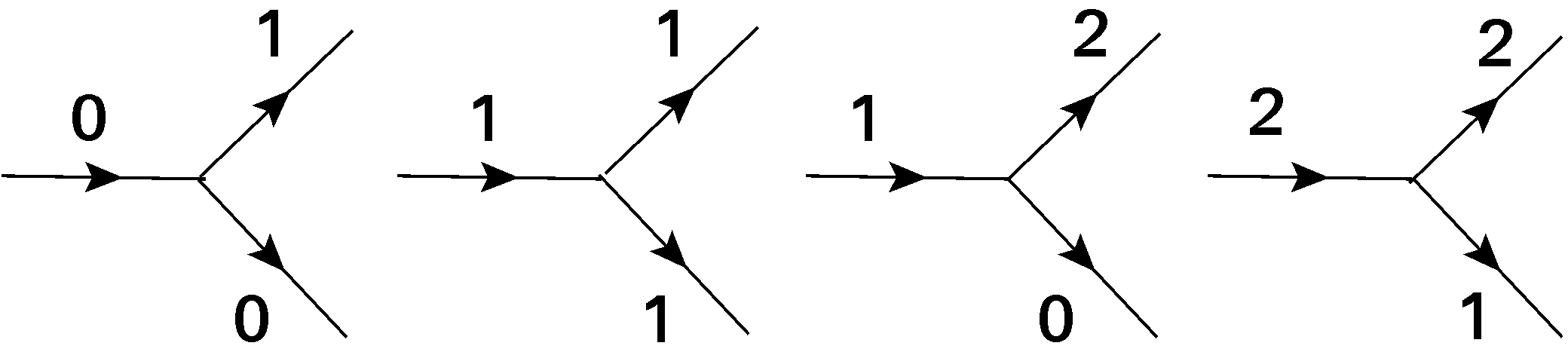}
\caption{\label{fig:3vertex} Relevant examples of 3-vertices of magnon decay processes are shown, along with the selection rules for magnetic quantum numbers.   }
\end{figure}

We observe that $\Gamma(\{m,\epsilon\})$ has a vanishing value at $\epsilon \to 0$. 
This statement becomes rather obvious when we represent the LSWT Hamiltonian in the form 
$H^{(2)} = \hat A^{\dagger}_{\mathbf{r}} \hat A_{\mathbf{r}} $ with $\hat A_{\mathbf{r}}$ given in \eqref{Hint} and $\hat A^{\dagger}_{\mathbf{r}}$ its Hermitian conjugate. \cite{Ivanov1999,Aristov2015}
To calculate the matrix elements of the interaction operators, we write  
\begin{equation}
\psi_{m,\epsilon \to 0} = c_{m,\epsilon} \frac{r^{m}}{1+r^{2}}\left(1-  \epsilon \,   {\cal O}(r^2)\right)
\label{Psi-low}
\end{equation}
 for $\epsilon \to 0$, $r\alt 1$, and   $c_{m,\epsilon}$ is  
the energy-dependent prefactor needed to provide the orthonormality of  the set $\Psi_{m,\epsilon}(\mathbf{r})$ in the definition  of the Green's function,  \eqref{Green}.

The quantity $c_{m,\epsilon}$ is discussed at some length in Appendix and consists of two factors. First factor is  the overall normalization  $\sim \epsilon^{1/4}$, which stems from the large distances and is the same for all $m$ in the thermodynamic limit;  together with the formula $\sum_{\epsilon} \to \int \epsilon^{-1/2}d \epsilon  $ this part leads to the usual density of states in two dimensions, $\rho(\epsilon) \sim 1$. The second factor in  $c_{m,\epsilon}$ is less trivial, as it refers to the skyrmion and arises from matching the short-distance form of $\psi$, Eq.\  \eqref{Psi-low}, with its far asymptote, $r\gg 1/\sqrt{\epsilon}$. 
This second factor is simply unity for two zero modes, $m=0$ and $m=2$, but is singular $\sim \epsilon^{-1/2} /\ln \epsilon$, for $m=1$.

We can write $ \hat A \Psi_{\epsilon \to 0,m}= \epsilon \,  c_{m,\epsilon} \frac{r^{1+m}}{1+r^{2}}   {\cal O}(1) $ and  
 estimate 
\begin{equation}
\Gamma(\{m,\epsilon\})= \frac{\epsilon}{\sqrt{s}} c_{m\epsilon} c_{m'\epsilon'}c_{m''\epsilon''} \int r\,dr \, \frac{r^{2+2m}}{(1+r^{2})^{4}} {\cal O}(1) , 
\label{F3estim}
\end{equation}
with $\epsilon$ corresponding to external energy  in Fig.\ \ref{SEdiagram}a ; the latter integral 
is convergent at $m < 2$, logarithmic divergence at $m=2$ is cut off by the largest of the energies.  We obtain the retarded self-energy part in the form  
\begin{equation}
\begin{aligned}
\Sigma(\omega,m, \epsilon) & \sim \frac{ \epsilon^2 \tilde \rho(m,\epsilon)  }{s} \int\frac{d\epsilon_1d\epsilon_2 \tilde\rho(m_{1},\epsilon_1)\tilde \rho(m_{2},\epsilon_2)}{\omega-(\epsilon_1+\epsilon_2)+i0} \,.
% \\  & \sim -  \frac{ \epsilon^2}{s} \left( \ln \frac1{|\omega|} + i \pi \, \theta(\omega)\right)
\end{aligned}
\label{16}
\end{equation}
When passing from summation in  \eqref{14} to integration here, $\sum_{\epsilon} \to \int \epsilon^{-1/2}d \epsilon$,  we defined the quantity  
\begin{equation}
\begin{aligned}
\tilde \rho(m,\epsilon) & =c_{m\epsilon}^{2}/\sqrt{\epsilon} \\ 
& \sim 1, \quad m=0,2 \,, \\ 
& \sim \epsilon^{-1} \ln^{-2} \epsilon, \quad m=1
\end{aligned}
\label{effDOS} 
\end{equation}
with integrable singularity in $\tilde \rho(1,\epsilon)$.  
Clearly, the principal contributions  to $\Sigma(\omega,m, \epsilon)$ are delivered by the internal lines with $m_{1,2}=1$, and the most troublesome contribution is given by the second vertex in Fig.\  \ref{fig:3vertex}. 
For the imaginary part we have at $\omega  \to0$

  \begin{equation}
\begin{aligned} 
\Im  \Sigma (\omega,m, \epsilon) 
&\sim    s^{-1}     { \epsilon^2}  \frac{\theta(\omega)}{\ln \omega} 
,  \quad  m=0,2  \, ,
\\
 &\sim s^{-1}    \frac { \epsilon} {\ln^{2}\epsilon}  \,    \frac{\theta(\omega)}{\omega \ln^{3} \omega} 
 , \quad  m=1  \, ,
\end{aligned}
\label{16a}
\end{equation}
with Heaviside function $\theta(\omega)=1$ at $\omega>0$. 
Analytic continuation is done according to the rule 
\begin{equation}
  \ln(-\omega) = \ln|\omega| - i\pi \theta(\omega)
\label{AnaCon}
\end{equation} 
which leads to the estimate of 
the most singular part of full expressions :
  \begin{equation}
\begin{aligned} 
 \Sigma(\omega,m, \epsilon) 
&\sim -    s^{-1} \epsilon^2    \ln( \ln (-1/\omega))   
,  \quad  m=0,2  \, ,
\\
 &\sim  s^{-1}    \frac { \epsilon} {\ln^{2}\epsilon}  \,    \frac{1}{\omega \ln^{2} (-\omega)} 
 , \quad  m=1  \, ,
\end{aligned}
\label{16b}
\end{equation}

% In the latter case the mode $m=1$ decays into two modes $m=1$, and we focus at this case now. 
  
The poles of the  Green's function  
\begin{equation}
\label{20}
G(\omega, m, \epsilon )= \left ({\omega-\epsilon-\Sigma(\omega,m,\epsilon)} \right)^{-1}
\end{equation}
define the renormalized magnon's energy. (We remind the reader that the 3-vertex is absent in case of uniform ferromagnetic state, and $\Sigma$ is strictly zero at $T=0$.)
We see from (\ref{16b}) that the relative corrections $\Sigma/\epsilon$ for low energies to the states $m=0,2$  are small and magnons remain well-defined quasiparticles.  At the same time, the second-order correction to $m=1$ state is not small and the calculation of higher order contributions is required.  This is achieved by dressing the 3-vertex, as discussed in the next section. 
 
 % \begin{figure} \includegraphics[scale=0.4]{5.png}
% \caption{Imaginary part of Self-Energy correction for two-dimensional and three-dimensional ferromagnet}
% \end{figure}

 %%%%%%%%%%%%%%%%%%%%%%%%%%%%%%
 %
 %%%%%%%%%%%%%%%%%%%%%%%%%%%%%%

\section{\label{sec:level4}Two-particle interaction and vertex corrections}
 
\begin{figure}
(a)\hspace{.2cm}\includegraphics[width=0.9\columnwidth]{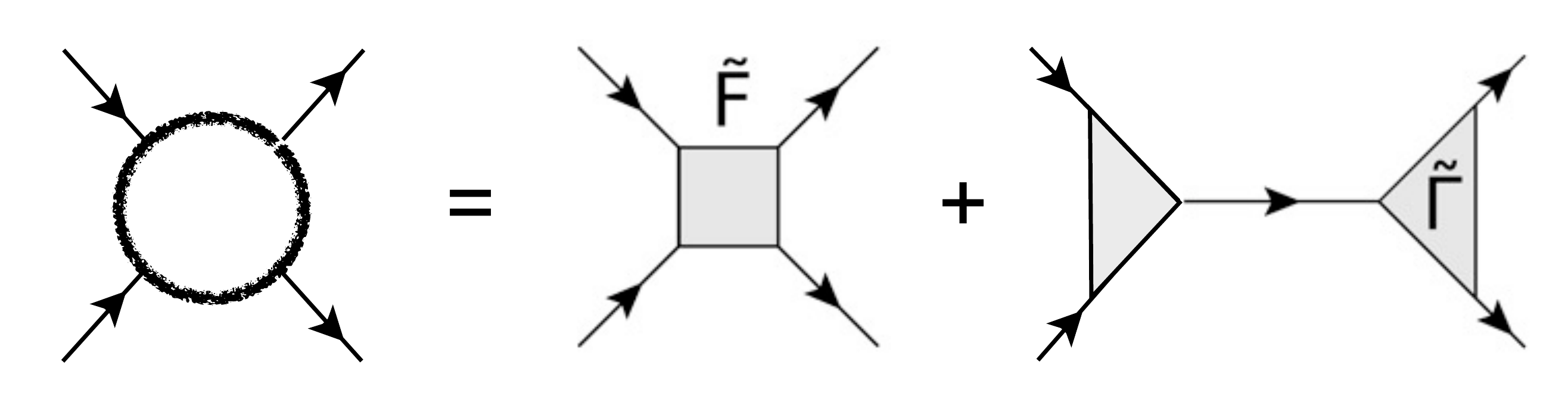}  \\
(b) \hspace{.2cm}\includegraphics[width=0.9\columnwidth]{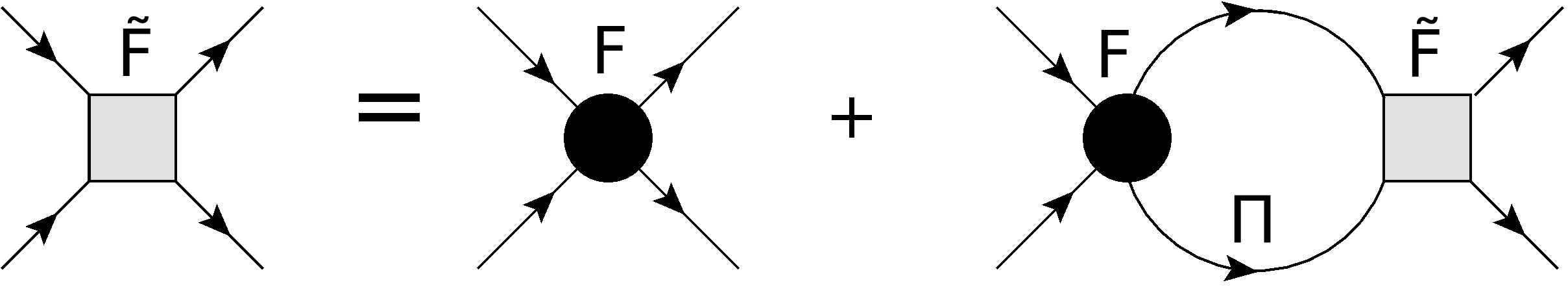} \\
(c)\hspace{.2cm}\includegraphics[width=0.9\columnwidth]{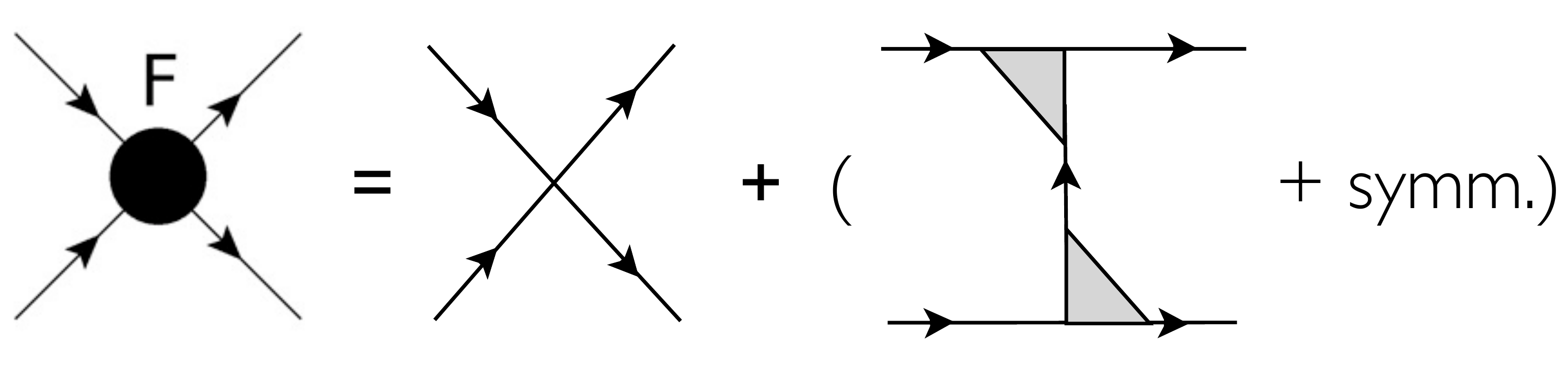}  \\
(d)\hspace{.2cm}\includegraphics[width=0.9\columnwidth]{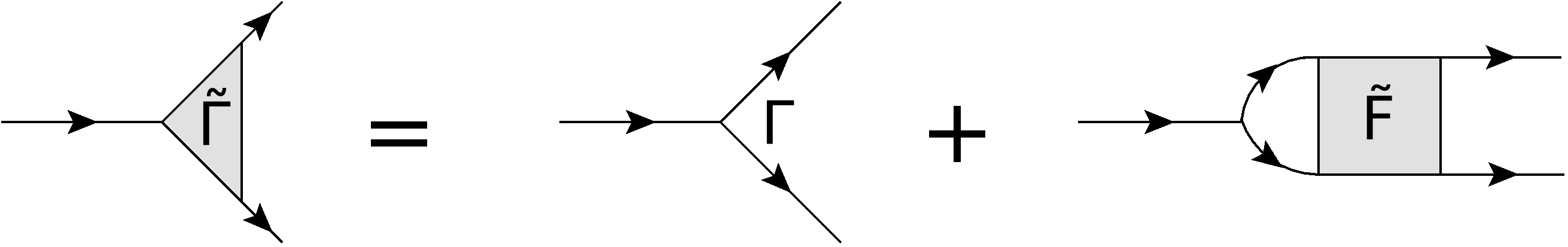}
\caption{\label{fig:vertices}
(a) Two particle Green's function represented as a sum of 1-irreducible and 1-reducible parts. 
(b) Bethe-Salpeter equation in graphical form for the 1-irreducible dressed vertex of interaction. 
(c) The ``bare'' part of the interaction as a sum of  the bare 4-vertex and the ladder rung diagram, the symmetrization over the external legs is assumed.
(d) The graphical equation for the  dressed 3-vertex,  shown as grey triangle in other graphs. }
\end{figure}

Let us now discuss the renormalization of the vertex  $ {H}_{int}^{(4)}$ in \eqref{Hint} which defines the two-particle Greens' function.  It is convenient to represent this function as the sum of two contributions, shown in Fig. \ref{fig:vertices}a . One of them, called 1-reducible part,  can be cut by one Green's function into two parts. The internal Green's functions should be dressed, but we saw above that the distinction between the dressed and the bare functions is inessential for $m\neq 1$.  For $m=1$ the situation is more delicate and we first proceed in this case with bare Green's functions and return to this point later.  

The 1-irreducible part can be represented by a sequence of diagrams, leading  to the Bethe-Salpeter (BS) equation, as shown in Fig. \ref{fig:vertices}b. In our case the bare scattering amplitude of two magnons is obtained in the order $1/s$ as a sum of two contributions shown in   Fig. \ref{fig:vertices}c. One contribution is merely ${H}_{int}^{(4)}$ and another contribution appears in presence of skyrmion and  is the rung element of ladder sequence, symmetrized over the external legs.  

The solution of the BS equation for the uniform FM ground state is characterized by poles in the renormalized vertex \cite{Wortis1963}. These poles signify the bound states of two magnons, occuring in any spatial dimension. For completeness we sketch   the derivation of this fact in 2D case. In the uniform FM state the spectrum is $\epsilon_{k} = k^{2}$ and the vertex is given by double gradient term in   $ {H}_{int}^{(4)}$, which can be formally obtained from  \eqref{Hint} by tending $r\to \infty$. Let  the total momentum of the incoming magnons be $2\mathbf{k}$.  We parametrize the internal momenta in the diagram Fig. \ref{fig:vertices}b as $\mathbf{k} \pm \mathbf{p}$ and arrive to the expression for the loop

\begin{equation}
\begin{aligned}
\Pi (\omega,2\mathbf{k}) &  = \int  \frac{d^{2} \mathbf{p} }{(2\pi)^{2}} \,\frac{p^{2}-k^{2}}{\omega- 2 k^{2} - 2 p^{2}} 
\\ & =- {\cal O}(1) + \frac{1}{4\pi} \left(k^{2} - \frac{\omega}4 \right) \ln \frac \Lambda{k^{2}-\omega/2}
\end{aligned} 
\label{eq:Loop}
\end{equation}
with $\Lambda \sim J $ ultraviolet cutoff. 
The latter expression shows that in the BS equation 
\begin{equation}
\label{26}
\frac1{2s} \nabla \nabla = F \rightarrow \tilde{F} = \frac{F}{1-F\Pi} = \frac{1}{2s-\Pi} \nabla \nabla 
\end{equation}
the denominator has poles at $\omega \simeq  2k^{2}  - \exp(-16\pi s/k^{2}) $, i.e. slightly below the minimum energy of two magnons with the total momentum $2\mathbf{k}$. Notice that the large logarithm in \eqref{eq:Loop} is obtained from integration over $p\gg k$, where the incoming momenta are almost antiparallel.
% and the amplitude of  $ (2s)^{-1} \nabla \nabla$ corresponds to attraction.

%\begin{figure} \includegraphics[width=0.9\columnwidth]{Renorm_2.png}
%\caption{The vertex of interaction} \end{figure}

%\begin{figure} \includegraphics[scale=0.09]{4.png} \caption{The vertex of interaction} \end{figure}

In case of the skyrmion  the situation is different.  We write at first the expression for the bare 4-amplitude 
with $\hat{H} ^{(4)}$ from \eqref{Hint}  : 
\begin{equation}
F(\{m,\epsilon\})=\frac1{2s} 
\int d^{2} \mathbf{r}\,    \Psi_{m_1,\epsilon_{1}}\Psi_{m_2,\epsilon_{2}}\hat{H} ^{(4)}\Psi^{*}_{m_3,\epsilon_{3}}\Psi^{*}_{m_4,\epsilon_{4}},   
\label{24}
\end{equation}
and notice that $F(\{m,\epsilon\})$ remains finite 
in the zero energy limit, $\epsilon_{i} \rightarrow 0$ for $m_{i}\neq 1$, whereas $F(\{m,\epsilon\})$ is singular when one of $m_{i}=1$.  We are mostly interested in this limit, in which case the wave functions are given by \eqref{Psi-low}, and focus on the most singular case, all $m_{i}=1$, which corresponds to second line in Eq.\ \eqref{16b}. (Less singular cases, e.g. $m_{3}=0$, $m_{4}=1$, correspond to the first line in  Eq.\ \eqref{16b} without factor $\epsilon^{2}$.) Part of the terms in $\hat{H} ^{(4)}$ effectively cancels by producing extra powers of $\epsilon_{i}$ when acting on the modes $\Psi^{*}_{m_3,\epsilon3} \Psi^{*}_{m_4,\epsilon4}$, except for the first term in the last line of Eqs.\ \eqref{Hint}. 
The integration by $\phi$ gives  $m_1+m_2=m_3+m_4 = M$ and apart from the factors $c_{m,\epsilon}$ in \eqref{Psi-low} we obtain 
\begin{equation}
 F(\{m,0\}) \sim \frac1{2s}  \int \frac{r^{2M+1}}{(r^2+1)^6} dr  >0
 \label{25}
\end{equation}
which is convergent at $0\le M\le 4$, i.e. for the desired range,  $0\le m_{i} \le 2$. 

The second type of diagrams in the combined vertex $F$ in Fig.\  \ref{fig:vertices}c, including the 3-vertices can be neglected in the considered limit of vanishing incoming energies. This statement is  verified for the bare quantity,  $\hat{H}^{(3)}$, which contains small energies of created or decaying magnon due to the operator $\hat A_{\mathbf{r}}$.  The existence of additional  energy factor in 3-vertex  shows  that the contribution of 1-reducible part in Fig. \ref{fig:vertices}a can also be neglected in the limit of vanishing energies.

The loop $\Pi$ for all $m_{i} =1$ is given by the expression 
\begin{equation}
\begin{aligned}
\Pi(\omega)  &\sim  \int\frac{d\epsilon_1d\epsilon_2 \tilde\rho(1,\epsilon_1)\tilde \rho(1,\epsilon_2)}{\omega-(\epsilon_1+\epsilon_2)+i0}  \, ,% \sim   \ln (- \omega )
\\ & 
 \sim    \frac{1}{\omega (\ln |\omega| - i\pi \theta(\omega)) ^{2}}  \,, 
\end{aligned}  \label{Pi1} 
\end{equation} 
in line with the estimates \eqref{16a}, \eqref{16b},  with $|\omega| \ll 1$ assumed in the derivation.   
The Bethe-Salpeter equation takes then the form 
\begin{equation}
\hat H^{(4)} \to 
 (1- F\Pi )^{-1} F  \sim \left ( s-\frac{1}{\omega \ln^{2} (-\omega) }  \,  {\cal O} (1)  \right)^{-1}
\label{BetheSal}
\end{equation}
with the constant $F$ given by \eqref{25}.  The obtained expression shows that in the dressed interaction vertex there are no poles at negative $\omega$, which corresponds to absence of the bound state. 
 At the same time,  for small positive total energy  of the incoming magnons, $\omega = \epsilon _{1} + \epsilon_{2}$, the dressed vertex \eqref{25} reveals a resonance, i.e. a pole in the complex plane of $\omega$.

It may be added that the above derivation is inapplicable for magnons with  $m\neq 0,1,2$, whose wave functions are suppressed at the center. The estimate \eqref{25} becomes invalid and one should apparently use the interaction terms \eqref{Hint} at large distances, $r\gg 1$, in which limit  $\hat H^{(3)}$ vanishes and $\hat H^{(4)}$ is reduced to the product of gradient terms. That means that far from the skyrmion the  picture of the usual ferromagnet is restored and we expect the existence of usual bound states.    The argument in favor of bound states away from the skyrmion also refers to the fact of exponentially small bound energy, which implies very large distances as compared with the skyrmion radius. 

Comparing  the usual bound state \eqref{26} with  
the resonance obtained  in \eqref{BetheSal}, we  observe that the usual bound state occurs at $\omega \simeq  2\epsilon  - \exp(-16\pi s/\epsilon) > 0$, i.e. above the vacuum energy, and below magnon continuum. The resonance  stemming from \eqref{BetheSal} happens for $s\gg1$ at 
\begin{equation}
\omega_{0} =  \frac{s^{-1}}{ \ln^{2}s} \left(1-i\frac{2\pi}{\ln s} \right)  {\cal O} (1)
\label{defW0}
\end{equation} 
which is slightly above the vacuum (classical) energy of a skyrmion configuration, Eq.\  \eqref{Eclass}; the resonance is well defined, $\mbox{Im }\omega_{0} \ll \mbox{Re }\omega_{0}$, for exponentially large $s$. Restoring the energy units and measuring the energy from the uniform ferromagnetic state, $E_{FM} =- s^2 J_{0} V$, we estimate 
$\omega_{0} = C ( 4\pi  s^{2} + { r_0^{-2}}{ } {\cal O} (\ln^{-2}s))$. This energy is higher for smaller $r_{0}$. 

The resonance lies within the magnon continuum and there is a possibility for magnons to decay to the resonance, which is absent in the uniform Heisenberg ferromagnet and possible in our situation. It is manifested as an additional pole at positive energies in the magnon Green's function. 
Consider the renormalization of the 3-vertex depicted in Fig.\ \ref{fig:vertices}d. Simple analysis shows that we can use the previous formulas for loop integration and write: 
\begin{equation}
\begin{aligned}
\tilde{\Gamma}& =\Gamma+\Gamma \Pi \tilde{F}=
\Gamma+\frac{\Gamma\Pi F}{1-F\Pi}=\frac{\Gamma}{1-F\Pi} , 
\end{aligned}
\end{equation}
This expression justifies our omittance of the ladder terms in Fig. \ref{fig:vertices}c and leads to  the improved estimate for the self energy :  
  \begin{equation}
\begin{aligned} 
 \Sigma(\omega,m=1, \epsilon) 
 &\sim    \frac { s^{-1}  \epsilon} {\ln^{2}\epsilon}  \,    \frac{1}{\omega \ln^{2} (-\omega) - {\cal O}(s^{-1})} 
 , 
\end{aligned}
\label{Sigma1}
\end{equation}
The last expression shows that, on one hand, we have better defined quasiparticles with $\omega \simeq \epsilon$ at $0 < \epsilon < |\omega_{0}|$. On the other hand, it also shows a resonance at $\omega \simeq \omega_{0}$ in the Green's function \eqref{20}. 
The residue at the resonance is dependent on $\epsilon$, has maximum at $\epsilon \simeq \omega_{0}$ and 
is logarithmically small for $\epsilon \gg \omega_{0}$ as $s^{-1} \ln^{-2}\epsilon \ln^{-2}\omega_{0}$.  
The damping at the resonance is largely independent of $\epsilon$, and thus defines the characteristic time scale of the skyrmion, $\tau^{-1} \sim \mbox{Im }\omega_{0}= C r_{0}^{-2} {\cal O}(\ln^{-3} s)$.  

We note that the appearance of resonance is the main distinction between the bare and the dressed Green's function, $G(\omega,m=1,\epsilon)$. This should lead to some modification of $\tilde\rho(1,\epsilon)$ in Eqs. \eqref{16}, \eqref{Pi1}.  The corresponding redistribution of the spectral weight around $\omega_{0}$ is smooth for moderate values of $s$ and can be neglected in our calculation.

%%%%%%%%%%%%%%%%%%%%%%%%%%
%  
%%%%%%%%%%%%%%%%%%%%%%%%%%  

Let us  now briefly discuss the generalization of our analysis for the 3D case. We conisder the layered structure with the ferromagnetic exchange interaction between layers of strength $0<J_{\perp}< J_{0} $. The low-energy spectrum for the uniform ferromagnetic case is $E   \simeq s C q^{2} + 2sJ_{\perp}(1 - \cos q_z )$. We assume that the  skyrmion configuration has the same form in plane as in 2D case and is independent of the third coordinate, $z$.  Eq. \eqref{Eclass} gives the skyrmion contribution to the classical energy $\delta E = 4\pi  C s^{2} L$,  with $L$ the system size in the third direction. The small-energy magnon spectrum   is  $E = \epsilon + C_{3} q_z^2$; if we measure energy in units of $ \epsilon_{0}$ and $q_{z}$ in inverse interlayer distances then $C_{3} \sim sJ_{\perp} / \epsilon_{0}$. The magnon wave function is now multiplied by the plane wave in the third direction, $\exp (i zq_{z})$. 
The vertices  $ {H}^{(3)}$,  $ {H}^{(4)}$  include the integration over the third coordinate. 
The vertex $\hat{H}^{(3)}$ in \eqref{Hint} remains the same, and   $\hat{H} ^{(4)}$  has also the gradient terms in the third direction, $ \nabla_{\mathbf{r}_{1}}\nabla_{\mathbf{r}} \to 
 \nabla_{\mathbf{r}_{1}}\nabla_{\mathbf{r}} +  (J_{\perp}/J) \,  \nabla_{z_{1}}\nabla_{z}$. 
 The most important part of the loop diagram with all $m_{i}=1$ now depends on the total momentum of incoming magnons in the third direction, $k_{z}$, and 
 takes the form, cf.\ \eqref{Pi1}   :  
\begin{equation}
\begin{aligned}
\Pi(\omega , k_{z})  & \simeq   \int_{0}^{{\cal O}(1)} dq_{z} \int\frac{d\epsilon_1d\epsilon_2  \, \rho(1,\epsilon_1)\rho(1,\epsilon_2) }{\bar{\omega}-\epsilon_1-\epsilon_2 + i0}  \\ & 
\simeq 
\int  \frac{dq_{z} }{\bar{\omega} \ln^{2}(-\bar \omega)}  \\
\end{aligned} \label{Pi3D0}
\end{equation}
with $\bar{\omega}=\omega- C_{3} k_z^2/2 - 2C_{3} q_{z}^2$, and the expression for the last integrand is obtained for $|\bar{\omega}| \alt 1$. In particular case, $k_z =0$, we  estimate the most singular contribution as 
\begin{equation}
\begin{aligned}
\Pi(\omega , k_{z}=0)  & \sim    \frac{1}{ {\omega} \ln^{2}(-  \omega)} , \quad  |\omega| \gg C_{3} \\
 & \sim    \frac{-1}{ \sqrt{-C_{3}\omega} \ln^{2}(-  \omega)} , \quad  |\omega| \alt C_{3} \\
\end{aligned} \label{Pi3D}
\end{equation}
For  $J_{\perp} \to 0$ the expression \eqref{16} is restored.    The crossover to 3D regime happens 
for   $C_{3}\agt \omega_{0}$, i.e. at 
\begin{equation}
 J_{\perp} \agt J_{0}  r_{0}^{-2} s^{-1} \ln^{-2}s .
 \label{crossover} 
\end{equation} 
In the latter case the value of $\Pi$ becomes almost entirely imaginary, so that the resonance is not formed. The dressing of the 3-vertex $\Gamma$ in this case leads to the obviously modified expression \eqref{Sigma1}, which shows well-defined magnons.   
The characteristic timescale in this fully developed 3D case may be estimated from the relation $|\Pi| \sim s^{-1}$ which yields $\tau^{-1} \sim \omega_{0}^{2} / C_{3}$.

 %%%%%%%%%%%%%%%%%%%%%%%%%%%%%%
 %
 %%%%%%%%%%%%%%%%%%%%%%%%%%%%%%

\section{\label{sec:level6} Discussion and  conclusions}
 
We discussed the stability of Belavin-Polyakov skyrmion in the Heisenberg ferromagnet for zero temperature. 
Such skyrmion is higher in energy than the usual uniform ferromagnet,  and we discuss whether  interaction between  magnons can produce instability of this metastable state.  The self-energy corrections due to magnon decay are calculated in 2D and 3D case, revealing no sizable effects in the small energy limit for all modes, except for the dilational mode, $m=1$. The latter soft mode corresponds to the dilatation and change of orientation of the skyrmion, and is characterized by vanishing energy. The crucial difference of this mode from other soft modes is the singular weight, $\sim  \epsilon^{-1/2}$,  which characterized  its wave-function at smaller distances and small energy, $\epsilon \ll s  J_{0}r_{0}^{-2}$.

This anomalous weight stems from the necessity of normalization of eigenfunctions in the thermodynamic limit and can be understood as ``tsunami effect'. Namely,  a spin wave starting at the infinity with a small amplitude and small energy (and thus a small velocity $\sim \sqrt{\epsilon}$) is enormously enhanced when traveling close to the skyrmion core. Such hypersensitivity of the breathing mode, $m=1$, results in the  eventual formation of the well defined resonance state of two magnons with $m=1$ in two dimensions. This resonance is observed in the solution of  Bethe-Salpeter equation and is also manifested as an additional pole in the one-magnon Green's function for the breathing mode.   In three dimensional case such resonance is not formed and we show a crossover scale between two regimes.
 
Our analysis suggests that the two particle Green's function, which demonstrated bound states in the usual ferromagnet, shows different behavior in presence of the skyrmion. In addition to formation of the resonance mode, the  Bethe-Salpeter equation in 2D shows that the magnon scattering is logarithmically suppressed near the skyrmion in higher orders of semiclassical parameter, $1/s$. No pole is found in the corresponding renormalized scattering amplitude.  The latter property does not exclude the formation of bound states at the inifinity, whose wave-functon is hence non-uniform and avoids the vicinity of the skyrmion.   
  
Our results were obtained for the special case of highly symmetric Hamiltonian and in the  limit of large spin $s$. For this Hamiltonian and in the zero temperature limit the considered processes exhaust all possible contributions to spin dynamics and it means the results are qualitatively valid also for realistic spin values $s\sim 1$.   

Consideration of  systems of lower spin symmetry, i.e.\  in the presence of Dzyaloshinskii-Moriya interaction and magnetic field, may modify our findings in several aspects. First of all, these additional interactions may lower the classical energy of the system in certain range of parameters. The quantum spectrum is now determined by the quadratic Hamiltonian, containing the anomalous terms in creation and annihilation operators and requiring an extended basis \cite{Schutte2014}. One can diagonalize this Hamiltonian by appropriate canonical transformation and come to proper magnon operators. The spectrum contains one   zero mode, corresponding to the broken translational symmetry, while the previous dilatational and SCT modes acquire finite energy.  One expects that the interaction vertices are much more complicated due to anomalous terms in quadratic Hamiltonian and necessity to operate in the extended basis.  The calculated corrections to zero mode should however be absent by symmetry,  and we expect the technical reason for this property in a special form of the decay 3-vertex amplitude, which will contain a half of spectral operator,  $\hat A_{\mathbf{r}}$, see Eq. \eqref{Hint}.  Our results about resonance in the dressed 4-vertex for dilational modes shall qualitatively be valid in the presence of additional interactions. Depending on relations between these interactions, one can suggest either a resonance or a true bound state below the continuum spectrum, which point requires further theoretical studies.   

\acknowledgements

We thank A.O. Sorokin, K.L. Metlov,  M. Garst,  A.S. Ovchinnikov, B.A. Ivanov, S.V. Maleyev, P. W\"olfle for useful discussions and communications. 
This work was supported by the Russian Scientific Foundation grant (project 14-22-00281).

\appendix*

\section{\label{app} Density of states  in the presence of BP skyrmion}

%  It is worthy to note, that the corresponding Shr\"{o}dinger equation is exactly solvable only for case of zero energy of a magnon ($\epsilon=0$, or for zero modes) and for $m=0$. However it is possible to find $\Psi_{m,\epsilon}(r)$ as a perturbation for zero modes, because in case $T=0$ only low-energy excitations provide the main contribution to Green's function and thermodynamical values  (see Appendix A).  

% (we place a skyrmion into an arbitrary large cylinder and replace a summation over states by an integration over energies, see also Appendix )

% \setcounter{equation}{0} \numberwithin{equation}{section}
We sketch the derivation of corrections to the density of states (DoS) in the presence of the skyrmion.   Consider the  eigenfunctions $\Psi_{m,\epsilon}(r)$ for the Hamiltonian \eqref{LSWT}. For $\epsilon =0$ we have Eq. \eqref{zeromodes} and for $\epsilon \ll 1$ we  write 
\begin{equation} 
\label{A1}
\Psi_{m,  \epsilon}=\Psi_{m,0}+\epsilon\Psi^{(1)}_m + {\cal O}(\epsilon^{2})
\end{equation}  
and  further $ \hat H^{(2)} \Psi^{(1)}_m = \Psi_{m,0} $.  The solution of this equation in general case 
of BP skyrmion with $Q\ge 1$ is discussed at length in \cite{Ivanov1999}.  It was shown there that the most significant correction to DoS comes from the states with $m=1$ (in our notation). We have explicitly
\[
\Psi^{(1)}_1  = \frac{ \Psi_{1,0} }4
\left( \frac{ 1- r^4}{r^{2}} \ln(1+r^{2}) +2r^{2} +2 \mbox{Li}_{2}(-r^{2})-1
\right)  
\]
with $\mbox{Li}_{2}(x)$ polylogarithm function. At $r \gg1 $ we have 
\begin{equation}
\begin{aligned}
\frac{\Psi_{1,  \epsilon}}{\epsilon }
&\simeq \frac1{\epsilon r} + \frac{r }2 (1-\ln r) - \frac{\ln^{2}r} r {\cal  O}(1)
\end{aligned}
\label{A2}
\end{equation}
Now we notice that for $r \gg1 $ the Hamiltonian $\hat H^{(2)}$ describes  free motion far from the skyrmion. The corresponding function is expressed in terms of Bessel function $J_{m}(x)$ and $Y_{m}(x)$ as 
\begin{equation}
\begin{aligned}
 \Psi_{m, \epsilon} & \sim  J_{2-m}(k r) + \tan \delta  \cdot Y_{2-m}( k r)   
 \end{aligned}
\label{A3}
\end{equation}
here $k = \sqrt{\epsilon}$ and  $\delta$ is the $k$-dependent scattering phase shift.  For $\epsilon \to 0$ there is an interval $1 \ll r \ll 1/k$ where we can match \eqref{A2} and the short-distance asymptotic of Eq.\ \eqref{A3} : 
\begin{equation} 
 \frac{ \Psi_{1, \epsilon}}{k}  \sim   
- \frac{\pi r }{4} \cot \delta + \frac1{k^{2} r} - \frac r2 \ln \frac{kr}2 + \frac r4 (1-2\gamma) 
\label{A4}
\end{equation}
with $\gamma \simeq 0.577$. This comparison gives 
\begin{equation}   \tan \delta = \frac {\pi}{2 \ln (c_{1}/k)} \ll 1\, , 
\label{A4a}
\end{equation}
with $c_{1}\simeq 0.681$, in agreement with \cite{Ivanov1999}. 
Now we need to place the skyrmion at the center of a disc of large radius $R \gg 1$. 
At the edge of the disc the wave function is set to  zero, and the $n$th eigenvalue of $\hat H^{(2)}$ with this boundary condition corresponds to $n$th zero of the Bessel functions in Eq.\ \eqref{A3}. We have at large $kR \gg 1$ :   
\begin{equation} 
\begin{aligned}
 \Psi_{1, \epsilon} (R) &   \simeq \sqrt{\frac2{\pi kR} } \cos (k R - 3 \pi/4 - \delta)
\end{aligned}
\label{A5}
\end{equation}
which yields for   $n$th eigenvalue 
\begin{equation} 
 k R -  \delta \simeq \pi n
\label{A6}
\end{equation}
The summation over the quantum states, $n$, is eventually replaced by the integration over $\epsilon$ : 
\begin{equation}
\sum_{n} \to \int \frac{dn}{d\epsilon}  d\epsilon  
\label{A11}
\end{equation}
with the weight \[ \frac{dn}{d\epsilon}  = \frac{1}{2k }  \frac{dn}{dk} 
 = \frac{R}{2\pi k} \left (1 - \frac{1}{R} \frac{d\delta}{dk}  \right) .
  \] 
The overall large prefactor $R/k$ here is compensated in Green's functions \eqref{Green} by the factor $k/2R$ coming from the normalization of the wave function, $2\pi \int_{0}^{R}dr\,r  |\Psi_{m, \epsilon}|^{2}=1$, the last integral  in the limit $R\to \infty$ mostly contributed by the asymptote  \eqref{A3}. This compensation gives the constant density of states  (for fixed $m$) for the uniform ferromagnetic state, when one uses the asymptotic form of wave-functions, \eqref{A3}, in the definition \eqref{Green}. The correction, $\frac{1}{R} \frac{d\delta}{dk} $, is  small  as can be seen at $m=1$ from Eqs.\ \eqref{A4a}, \eqref{A6}  : 
 \[    \frac{1}{R} \frac{d\delta}{dk}  = \frac{\pi}{2kR} \frac{1}{\ln^{2}(c_{1}/k)} \ll 1.
  \] 
 Therefore we can use the same expression for the density of states, as in the case of the uniform ferromagnetic ground state. This conclusion is different from  \cite{Ivanov1999}, where the factor $1/R$ before ${d\delta}/{dk}$ was ignored.

At the same time, the form of the wave functions, appearing in the Green's function \emph{at smaller distances}, acquires the factor coming from the above matching of two asymptotes. If we use the constant density of states, as discussed above, then the matching coefficient  between \eqref{A2} and 
\eqref{A4} is  $\sim 1/k\ln(1/k) \gg 1$. It means that the form \eqref{zeromodes} is approximately valid for $r \sim 1$  for $\epsilon \neq 0$, but the prefactor, coming from the normalization at large distances, is huge for $m=1$. This is not so for other zero modes,  $m=0 , 2$, where this prefactor is unity. 

If we use Green's function \eqref{Green} with the orthonormal set of functions and want to use the form \eqref{zeromodes} in \eqref{F3estim}, \eqref{25}, then the discussed matching factors, being independent of $r$, can  be included into the definition of the Green's function in $(\omega,m,\epsilon)$ representation. After this redefinition we arrive to the \emph{effective} density of states  in the form of Eq.\ \eqref{effDOS}. 
 
% \bibliography{../../../Bibliography/MainBib.bib,../../../Bibliography/oldies.bib} \end{document}

\begin{thebibliography}{21}
\expandafter\ifx\csname natexlab\endcsname\relax\def\natexlab#1{#1}\fi
\expandafter\ifx\csname bibnamefont\endcsname\relax
  \def\bibnamefont#1{#1}\fi
\expandafter\ifx\csname bibfnamefont\endcsname\relax
  \def\bibfnamefont#1{#1}\fi
\expandafter\ifx\csname citenamefont\endcsname\relax
  \def\citenamefont#1{#1}\fi
\expandafter\ifx\csname url\endcsname\relax
  \def\url#1{\texttt{#1}}\fi
\expandafter\ifx\csname urlprefix\endcsname\relax\def\urlprefix{URL }\fi
\providecommand{\bibinfo}[2]{#2}
\providecommand{\eprint}[2][]{\url{#2}}

\bibitem[{\citenamefont{Fert et~al.}(2013)\citenamefont{Fert, Cros, and
  Sampaio}}]{Fert2013}
\bibinfo{author}{\bibfnamefont{A.}~\bibnamefont{Fert}},
  \bibinfo{author}{\bibfnamefont{V.}~\bibnamefont{Cros}}, \bibnamefont{and}
  \bibinfo{author}{\bibfnamefont{J.}~\bibnamefont{Sampaio}},
  \bibinfo{journal}{Nature Nanotechnology} \textbf{\bibinfo{volume}{8}},
  \bibinfo{pages}{152} (\bibinfo{year}{2013}).

\bibitem[{\citenamefont{Koshibae et~al.}(2015)\citenamefont{Koshibae, Kaneko,
  Iwasaki, Kawasaki, Tokura, and Nagaosa}}]{Koshibae2015}
\bibinfo{author}{\bibfnamefont{W.}~\bibnamefont{Koshibae}},
  \bibinfo{author}{\bibfnamefont{Y.}~\bibnamefont{Kaneko}},
  \bibinfo{author}{\bibfnamefont{J.}~\bibnamefont{Iwasaki}},
  \bibinfo{author}{\bibfnamefont{M.}~\bibnamefont{Kawasaki}},
  \bibinfo{author}{\bibfnamefont{Y.}~\bibnamefont{Tokura}}, \bibnamefont{and}
  \bibinfo{author}{\bibfnamefont{N.}~\bibnamefont{Nagaosa}},
  \bibinfo{journal}{Japanese Journal of Applied Physics}
  \textbf{\bibinfo{volume}{54}}, \bibinfo{pages}{053001}
  (\bibinfo{year}{2015}).

\bibitem[{\citenamefont{Romming et~al.}(2013)\citenamefont{Romming, Hanneken,
  Menzel, Bickel, Wolter, von Bergmann, Kubetzka, and
  Wiesendanger}}]{Romming2013}
\bibinfo{author}{\bibfnamefont{N.}~\bibnamefont{Romming}},
  \bibinfo{author}{\bibfnamefont{C.}~\bibnamefont{Hanneken}},
  \bibinfo{author}{\bibfnamefont{M.}~\bibnamefont{Menzel}},
  \bibinfo{author}{\bibfnamefont{J.~E.} \bibnamefont{Bickel}},
  \bibinfo{author}{\bibfnamefont{B.}~\bibnamefont{Wolter}},
  \bibinfo{author}{\bibfnamefont{K.}~\bibnamefont{von Bergmann}},
  \bibinfo{author}{\bibfnamefont{A.}~\bibnamefont{Kubetzka}}, \bibnamefont{and}
  \bibinfo{author}{\bibfnamefont{R.}~\bibnamefont{Wiesendanger}},
  \bibinfo{journal}{Science} \textbf{\bibinfo{volume}{341}},
  \bibinfo{pages}{636–639} (\bibinfo{year}{2013}).

\bibitem[{\citenamefont{Belavin and Polyakov}(1975)}]{Belavin1975}
\bibinfo{author}{\bibfnamefont{A.~A.} \bibnamefont{Belavin}} \bibnamefont{and}
  \bibinfo{author}{\bibfnamefont{A.~M.} \bibnamefont{Polyakov}},
  \bibinfo{journal}{JETP Lett.} \textbf{\bibinfo{volume}{22}},
  \bibinfo{pages}{245} (\bibinfo{year}{1975}).
  
  
\bibitem[{\citenamefont{Bogdanov and Hubert}(1994)}]{Bogdanov1994}
\bibinfo{author}{\bibfnamefont{A.}~\bibnamefont{Bogdanov}} \bibnamefont{and}
  \bibinfo{author}{\bibfnamefont{A.}~\bibnamefont{Hubert}},
  \bibinfo{journal}{Journal of Magnetism and Magnetic Materials}
  \textbf{\bibinfo{volume}{138}}, \bibinfo{pages}{255} (\bibinfo{year}{1994}).

\bibitem[{\citenamefont{Ezawa}(2010)}]{Ezawa2010}
\bibinfo{author}{\bibfnamefont{M.}~\bibnamefont{Ezawa}},
  \bibinfo{journal}{Phys. Rev. Lett.} \textbf{\bibinfo{volume}{105}},
  \bibinfo{pages}{197202} (\bibinfo{year}{2010}).

\bibitem[{\citenamefont{Mochizuki}(2012)}]{Mochizuki2012}
\bibinfo{author}{\bibfnamefont{M.}~\bibnamefont{Mochizuki}},
  \bibinfo{journal}{Phys. Rev. Lett.} \textbf{\bibinfo{volume}{108}},
  \bibinfo{pages}{017601} (\bibinfo{year}{2012}).

\bibitem[{\citenamefont{Han et~al.}(2010)\citenamefont{Han, Zang, Yang, Park,
  and Nagaosa}}]{Han2010}
\bibinfo{author}{\bibfnamefont{J.~H.} \bibnamefont{Han}},
  \bibinfo{author}{\bibfnamefont{J.}~\bibnamefont{Zang}},
  \bibinfo{author}{\bibfnamefont{Z.}~\bibnamefont{Yang}},
  \bibinfo{author}{\bibfnamefont{J.-H.} \bibnamefont{Park}}, \bibnamefont{and}
  \bibinfo{author}{\bibfnamefont{N.}~\bibnamefont{Nagaosa}},
  \bibinfo{journal}{Phys. Rev. B} \textbf{\bibinfo{volume}{82}},
  \bibinfo{pages}{094429} (\bibinfo{year}{2010}).


\bibitem[{\citenamefont{Nagaosa and Tokura}(2013)}]{Nagaosa2013}
\bibinfo{author}{\bibfnamefont{N.}~\bibnamefont{Nagaosa}} \bibnamefont{and}
  \bibinfo{author}{\bibfnamefont{Y.}~\bibnamefont{Tokura}},
  \bibinfo{journal}{Nature Nanotechnology} \textbf{\bibinfo{volume}{8}},
  \bibinfo{pages}{899} (\bibinfo{year}{2013}).
  
\bibitem[{\citenamefont{Aristov et~al.}(2015)\citenamefont{Aristov, Kravchenko,
  and Sorokin}}]{Aristov2015}
\bibinfo{author}{\bibfnamefont{D.}~\bibnamefont{Aristov}},
  \bibinfo{author}{\bibfnamefont{S.}~\bibnamefont{Kravchenko}},
  \bibnamefont{and} \bibinfo{author}{\bibfnamefont{A.}~\bibnamefont{Sorokin}},
  \bibinfo{journal}{JETP Letters} \textbf{\bibinfo{volume}{102}},
  \bibinfo{pages}{455} (\bibinfo{year}{2015}).


\bibitem[{\citenamefont{Yu et~al.}(2010)\citenamefont{Yu, Onose, Kanazawa,
  Park, Han, Matsui, Nagaosa, and Tokura}}]{Yu2010}
\bibinfo{author}{\bibfnamefont{X.~Z.} \bibnamefont{Yu}},
  \bibinfo{author}{\bibfnamefont{Y.}~\bibnamefont{Onose}},
  \bibinfo{author}{\bibfnamefont{N.}~\bibnamefont{Kanazawa}},
  \bibinfo{author}{\bibfnamefont{J.~H.} \bibnamefont{Park}},
  \bibinfo{author}{\bibfnamefont{J.~H.} \bibnamefont{Han}},
  \bibinfo{author}{\bibfnamefont{Y.}~\bibnamefont{Matsui}},
  \bibinfo{author}{\bibfnamefont{N.}~\bibnamefont{Nagaosa}}, \bibnamefont{and}
  \bibinfo{author}{\bibfnamefont{Y.}~\bibnamefont{Tokura}},
  \bibinfo{journal}{Nature} \textbf{\bibinfo{volume}{465}},
  \bibinfo{pages}{901} (\bibinfo{year}{2010}).

\bibitem[{\citenamefont{Iwasaki et~al.}(2014)\citenamefont{Iwasaki, Beekman,
  and Nagaosa}}]{Iwasaki2014}
\bibinfo{author}{\bibfnamefont{J.}~\bibnamefont{Iwasaki}},
  \bibinfo{author}{\bibfnamefont{A.~J.} \bibnamefont{Beekman}},
  \bibnamefont{and} \bibinfo{author}{\bibfnamefont{N.}~\bibnamefont{Nagaosa}},
  \bibinfo{journal}{Phys. Rev. B} \textbf{\bibinfo{volume}{89}},
  \bibinfo{pages}{064412} (\bibinfo{year}{2014}).

\bibitem[{\citenamefont{Lin et~al.}(2014)\citenamefont{Lin, Batista, and
  Saxena}}]{Lin2014}
\bibinfo{author}{\bibfnamefont{S.-Z.} \bibnamefont{Lin}},
  \bibinfo{author}{\bibfnamefont{C.~D.} \bibnamefont{Batista}},
  \bibnamefont{and} \bibinfo{author}{\bibfnamefont{A.}~\bibnamefont{Saxena}},
  \bibinfo{journal}{Phys. Rev. B} \textbf{\bibinfo{volume}{89}},
  \bibinfo{pages}{024415} (\bibinfo{year}{2014}).

\bibitem[{\citenamefont{Sch\"utte and Garst}(2014)}]{Schutte2014}
\bibinfo{author}{\bibfnamefont{C.}~\bibnamefont{Sch\"utte}} \bibnamefont{and}
  \bibinfo{author}{\bibfnamefont{M.}~\bibnamefont{Garst}},
  \bibinfo{journal}{Phys. Rev. B} \textbf{\bibinfo{volume}{90}},
  \bibinfo{pages}{094423} (\bibinfo{year}{2014}).

\bibitem[{\citenamefont{Tatara and Fukuyama}(2014)}]{Tatara2014}
\bibinfo{author}{\bibfnamefont{G.}~\bibnamefont{Tatara}} \bibnamefont{and}
  \bibinfo{author}{\bibfnamefont{H.}~\bibnamefont{Fukuyama}},
  \bibinfo{journal}{J. Phys. Soc. Jpn.} \textbf{\bibinfo{volume}{83}},
  \bibinfo{pages}{104711} (\bibinfo{year}{2014}).

\bibitem[{\citenamefont{Kovrizhin et~al.}(2013)\citenamefont{Kovrizhin,
  Dou\ifmmode~\mbox{\c{c}}\else \c{c}\fi{}ot, and Moessner}}]{Kovrizhin2013}
\bibinfo{author}{\bibfnamefont{D.~L.} \bibnamefont{Kovrizhin}},
  \bibinfo{author}{\bibfnamefont{B.}~\bibnamefont{Dou\ifmmode~\mbox{\c{c}}\else
  \c{c}\fi{}ot}}, \bibnamefont{and}
  \bibinfo{author}{\bibfnamefont{R.}~\bibnamefont{Moessner}},
  \bibinfo{journal}{Phys. Rev. Lett.} \textbf{\bibinfo{volume}{110}},
  \bibinfo{pages}{186802} (\bibinfo{year}{2013}).

\bibitem[{\citenamefont{Verga}(2014)}]{Verga2014}
\bibinfo{author}{\bibfnamefont{A.~D.} \bibnamefont{Verga}},
  \bibinfo{journal}{Phys. Rev. B} \textbf{\bibinfo{volume}{90}},
  \bibinfo{pages}{174428} (\bibinfo{year}{2014}).

\bibitem[{\citenamefont{Buijnsters et~al.}(2014)\citenamefont{Buijnsters,
  Fasolino, and Katsnelson}}]{Buijnsters2014}
\bibinfo{author}{\bibfnamefont{F.~J.} \bibnamefont{Buijnsters}},
  \bibinfo{author}{\bibfnamefont{A.}~\bibnamefont{Fasolino}}, \bibnamefont{and}
  \bibinfo{author}{\bibfnamefont{M.~I.} \bibnamefont{Katsnelson}},
  \bibinfo{journal}{Phys. Rev. B} \textbf{\bibinfo{volume}{89}},
  \bibinfo{pages}{174433} (\bibinfo{year}{2014}).

\bibitem[{\citenamefont{Ivanov and Stephanovich}(1989)}]{Ivanov1989}
\bibinfo{author}{\bibfnamefont{B.}~\bibnamefont{Ivanov}} \bibnamefont{and}
  \bibinfo{author}{\bibfnamefont{V.}~\bibnamefont{Stephanovich}},
  \bibinfo{journal}{Physics Letters A} \textbf{\bibinfo{volume}{141}},
  \bibinfo{pages}{89–94} (\bibinfo{year}{1989}).

\bibitem[{\citenamefont{Ivanov}(1995)}]{Ivanov1995}
\bibinfo{author}{\bibfnamefont{B.~A.} \bibnamefont{Ivanov}},
  \bibinfo{journal}{JETP Letters} \textbf{\bibinfo{volume}{61}},
  \bibinfo{pages}{917} (\bibinfo{year}{1995}).

\bibitem[{\citenamefont{Ivanov et~al.}(1999)\citenamefont{Ivanov, Murav'ev, and
  Sheka}}]{Ivanov1999}
\bibinfo{author}{\bibfnamefont{B.~A.} \bibnamefont{Ivanov}},
  \bibinfo{author}{\bibfnamefont{V.~M.} \bibnamefont{Murav'ev}},
  \bibnamefont{and} \bibinfo{author}{\bibfnamefont{D.~D.} \bibnamefont{Sheka}},
  \bibinfo{journal}{Journal of Experimental and Theoretical Physics}
  \textbf{\bibinfo{volume}{89}}, \bibinfo{pages}{583} (\bibinfo{year}{1999}).

\bibitem[{\citenamefont{Wortis}(1963)}]{Wortis1963}
\bibinfo{author}{\bibfnamefont{M.}~\bibnamefont{Wortis}},
  \bibinfo{journal}{Phys. Rev.} \textbf{\bibinfo{volume}{132}},
  \bibinfo{pages}{85} (\bibinfo{year}{1963}).

\end{thebibliography}

\end{document}